\documentclass[a4paper]{JAC2003}

\usepackage{graphicx}
\usepackage{amsmath}
\usepackage{textcomp}
\usepackage[]{caption}
\usepackage{subcaption}
\usepackage{booktabs}
\usepackage{multirow}
\usepackage{multicol}
\usepackage{varwidth}
\usepackage{xcolor}

\begin{document}
\title{OBSERVATIONS OF BEAM--BEAM EFFECTS AT THE LHC}

\author{G. Papotti, X. Buffat, W. Herr, R. Giachino, T. Pieloni  \\CERN, Geneva, Switzerland}

\maketitle

\begin{abstract}

This paper introduces a list of observations related to the beam--beam interaction that were collected over the first years of LHC proton physics operation (2010--12).
Beam--beam related effects not only have been extensively observed and recorded, but have also shaped the operation of the LHC for high-intensity proton running in a number of ways: the construction of the filling scheme, the choice of luminosity levelling techniques, measures to mitigate instabilities, and the choice of settings for improving performance (e.g.\ to reduce losses), among others.

\end{abstract}

\section{Introduction}

The Large Hadron Collider (LHC) at CERN, Geneva, is a 27\,km long circular accelerator~\cite{bib:LHC}.
It is based on a superconducting two-in-one magnet design, with dipoles that  allow it to reach a design energy of 7\,TeV per beam.
It features eight straight sections: four Interaction Points (IPs) are reserved for accelerator equipment and four house particle physics experiments.
IP3 and IP7 are dedicated to the collimation system, IP4 houses the RF system and most of the beam instrumentation, while IP6 is reserved for the beam dump system.
IP1 and IP5 contain the high-luminosity experiments ATLAS and CMS, while IP2 and IP8 accommodate the Alice and LHCb experiments, together with beam injection (Beam 1 through IP2, clockwise; Beam 2 through IP8, counterclockwise).

The luminosity requirements of the four experiments are very different~\cite{bib:Jaco}.
Two high-luminosity experiments and the discovery of a new boson are the reason for the push towards high-intensity proton physics performance.
This is detailed in the next section, where the beam parameters are compared between the Design Report and 2012 operation.

Alice and LHCb have luminosity limitations, and thus techniques of luminosity levelling have been applied consistently during proton physics production and will be described next.
The different luminosity requirements also impact the construction of the filling schemes.
Various collision patterns have been used for physics production and during 2012 a change was required to overcome recurrent loss of Landau damping.

The beam parameters were pushed much further during single-bunch Machine Development (MD) sessions, achieving very high beam--beam tune shifts.
Similar conditions were used for high pile-up studies for the experiments~\cite{bib:Trad}.

Scans of the crossing angle were done during MD sessions to evaluate the effect of long-range interactions in bunch trains, allowing the onset of losses for scaling laws to be measured~\cite{bib:Herr}.
The description of these studies and of the observation of orbit effects conclude this paper.

\section{Beam parameters and performance}

In these first years of luminosity production, the operation of the LHC has exceeded all expectations.
The year 2010 was mostly a commissioning year, and the instantaneous luminosity target was exceeded by a factor of 2, as $2.1\times10^{32}$\,cm$^{-2}\cdot$s$^{-1}$ was achieved.
The years 2011 and 2012 were dedicated to luminosity production in search for new physics, and 5.5 fb$^{-1}$
and 23.2 fb$^{-1}$ were collected in each year, respectively.
Table~\ref{tab:params} shows the machine and beam parameters as defined in~\cite{bib:LHC} compared to the ones used in 2012 operation.
Despite the beam energy being about half the design value, the achieved peak luminosity was over 75\% of the design value of $10^{34}$\,cm$^{-2}\cdot$s$^{-1}$.
The $\beta^{*}$ at the high luminosity experiments in IP1 and IP5 almost reached design values thanks to the excellent physical aperture and the use of `tight collimators'~\cite{bib:Rod}.

\begin{table}[b]
	\centering
	\caption{A Comparison of Parameters between Design Values~\cite{bib:LHC} and What Was Achieved in 2012 Operation}
	\begin{tabular}{lcc}
		\toprule{\textbf{Parameter}}					&{\textbf{Design}}&{\textbf{2012}}\\
		\toprule
			Beam injection energy [TeV]				& 0.45	& 0.45	\\  \midrule
			Beam energy at collisions [TeV]			& 7 		& 4 \\  \midrule
			Number of bunches					& 2808 	& 1380\\  \midrule
			Bunch spacing [ns] 					& 25 		& 50 \\  \midrule
			$\beta^{*}$ [m] 						& 0.55	& 0.6 \\  \midrule
			Intensity [$10^{11}$\,p/bunch]				& 1.15	& 1.65	\\  \midrule
			Norm.\ transv.\ emittance [$\mu$m]		& 3.75	& 2.5	\\  \midrule
			Beam size [$\mu$m]					& 16		& 19	\\  \midrule
			Peak luminosity [$10^{34}$\,cm$^{-2}\cdot$s$^{-1}$]	& 1	& 0.77	\\  \midrule
			Stored energy [MJ]					& 362	& 145	\\
		\bottomrule
	\end{tabular}
	\label{tab:params}
\end{table}

The key ingredient in the excellent luminosity performance is the fact that the LHC injectors can deliver much brighter beams with a bunch spacing of 50\,ns compared to the nominal 25\,ns.
At 4\,TeV beam energy, the pile-up $\mu$ (the number of inelastic collisions per bunch crossing) is at most 30--35, and this is still acceptable for the high-luminosity experiments.
This contributed to the choice, for 2012 operation, of 50\,ns spaced beams, which have the additional advantage of being much less affected by electron cloud than 25\,ns spaced beams (this allowed less beam time to be sacrificed to electron-cloud scrubbing, as 3 days were needed for 50\,ns versus the 2 weeks that would have been required for 25\,ns).
Note also that the smaller emittance of the 50\,ns beams allowed squeezing to proceed further (for comparison, $\beta_{25}^{*}=80$\,cm) and the use of a smaller crossing angle, both of which contributed directly to the excellent performance.

For operation after the Long Shutdown of 2013--14 (LS1), the pile-up $\mu$ will increase due to the energy increase, and thus 25\,ns is the preferred choice (for 50\,ns beams, $\mu_{50}\approx80$--120).
It is worth pointing out that luminosity levelling techniques might be needed even with 25\,ns spaced bunches as $\mu_{25}\approx25$--45.

\section{Luminosity levelling}

The Alice and LHCb experiments run with strong pile-up limitations: Alice at $\mu\approx0.02$ and LHCb at $\mu\approx2.5$.
The limitations come from various factors that range from detector damage through event size limitations to data-taking optimization~\cite{bib:Jaco}.
In addition to a less aggressive $\beta^{*}$ (in 2012, $\beta^{*}=3$\,m was used for IP2 and IP8), various techniques of luminosity control and levelling have been used operationally or tested in special runs at the LHC so far.

The luminosity was levelled operationally at LHCb so that the experiment could run at a constant luminosity of $4\times10^{32}$\,cm$^{-2}\cdot$s$^{-1}$. 
This was achieved by transversely offsetting the beams at the IP.
During the fill, the offset was adjusted in small steps so to modulate the overlap between the two beams to obtain the desired rates~\cite{bib:Jacq}.
No real limitations to this technique were found, as long as the offset bunch pair had enough tune spread due to head-on collisions elsewhere (i.e.\ in IP1 and IP5).

Given that the limitations in Alice are even stronger, the experiment ran for most of 2012 based on collision with so-called `satellite' bunches (`main-satellite' collisions).
Satellite bunches have a much lower charge (about a factor of 1000 lower than the main bunches), contained in buckets at 25\,ns from the main ones (which are at a 50\,ns spacing).
Note that this technique is not applicable with 25\,ns spaced bunches.

During MD sessions, techniques for $\beta^{*}$ levelling were also tested, verifying the feasibility and quality of the orbit control while squeezing IP1 and IP5. 
The squeeze of IP1 and IP5 was done in steps until the operational value of  60\,cm~\cite{bib:Wen1,bib:Wen2,bib:Mura}.

\begin{figure}[t]
\centering
\includegraphics[trim=0cm 5.3cm 0cm 0cm, clip=true, angle=0, width=75mm, height=40mm]{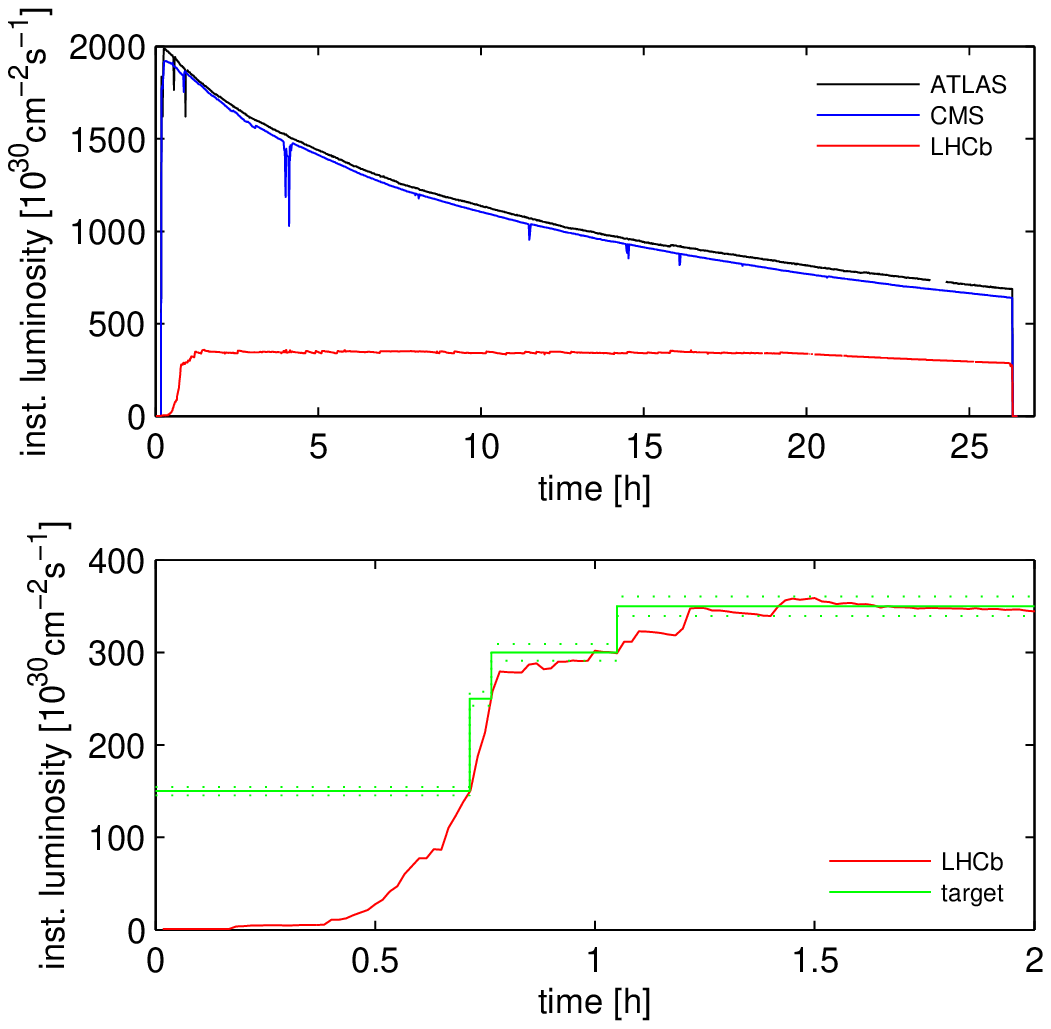}
\caption{Examples of ATLAS, CMS, and LHCb instantaneous luminosities during fill 2006 (in 2011). Note that while the luminosities of IP1/5 decrease with time, the luminosity of IP8 is kept constant by reducing the transverse offset and starts to decay only after about 20 h into production.}
\label{fig:nLRx}
\end{figure}

\section{Filling schemes and collision patterns}

Here, we recall a few of the constraints that have to be taken into account in the creation of a filling scheme:
\begin{itemize}

\item \emph{Experiment location}: ATLAS, Alice, CMS are located at the IP symmetry point, while LHCb is 11.25\,m away from it; ATLAS and CMS are diametrically opposed.

\item \emph{Kicker gaps}: the injection and extraction kickers require part of the ring not to contain beam (e.g.\ 925\,ns for the LHC injection kicker, and 3000\,ns for the dump kicker).

\item \emph{The 400\,MHz RF system}: this gives 2.5\,ns long buckets and a harmonic number $h=35\,640$, but 25\,ns bunch spacing is the minimum that the experiments' readout can handle (for a maximum of $\approx2800$ bunches per ring, taking the kicker gaps into account).

\item \emph{Bunch spacings}: the spacings that can be created in the LHC injector chain are 25\,ns, 50\,ns, 75\,ns, 150\,ns, or $>$250\,ns.

\item \emph{PS batch injections}: the number of injections into the SPS can be varied dynamically (i.e. from one to four injections).

\end{itemize}

Different numbers of colliding pairs are provided to the different experiments by shifting the injection buckets appropriately.
In Table~\ref{tab:schemes}, three examples of filling schemes used in 2012 for physics production are shown.
All three schemes are based on 50\,ns spaced bunches and main-satellite collisions for Alice (thus zero main--main collisions in IP2).
The first scheme was the baseline for 2012 operation, and it aimed at giving the same number of colliding pairs to IP1/5 and IP8.
Scheme 2 was designed to have all bunches colliding in IP1 and IP5, and was obtained by shifting four injections in scheme 1.
Scheme 3 is a minor modification with respect to scheme 2, designed to include three bunches with no collisions in IP1/5 for systematic background studies for ATLAS and CMS.

\begin{table}[t]
	\centering
	\caption{The Numbers of Collisions per IP for Three Filling Schemes Used in 2012}
	\begin{tabular}{lccc}
		\toprule{\textbf{Scheme}}	&{\textbf{IP1/5}}		&{\textbf{IP2}}	&{\textbf{IP8}}\\
		\toprule
			1				& 1331			& 0			& 1320\\  \midrule
			2				& 1380			& 0			& 1274\\  \midrule
			3				& 1377			& 0			& 1274\\
		\bottomrule
	\end{tabular}
	\label{tab:schemes}
\end{table}


\subsection{Loss of Landau Damping}

The change from scheme 1 to scheme 2 in Table~\ref{tab:schemes} was dictated by the fact that fills were often terminated prematurely due to instabilities.
Some bunches in ring 1 were losing intensity very quickly and an interlock kicked in at $\approx4\times10^{10}$\,p/bunch, effectively determining that the length of the fill was much shorter than desirable.

The affected bunches had the peculiarity of colliding only in IP8 (levelled by separation).
The lack of Landau damping with respect to the other bunches that collide in IP1/5 was identified to be the reason for the development of the instability~\cite{bib:Buf1}.
The filling scheme was thus changed to have head-on collisions in IP1/5 for all bunches, so that the head-on beam--beam tune spread would provide the necessary damping.

During the second part of the 2012 run, selected bunches frequently became unstable at the end of the squeeze, before collisions.
The instability was visible on loss measurements and as emittance growth, but it is not yet fully understood at the time of writing and studies are still ongoing~\cite{bib:Buf1}.
Improvements in beam instrumentation, and in particular for the detection of instabilities, are needed~\cite{bib:Giac}: for example, calibrated bunch-by-bunch emittance measurements, headtail monitors to understand the intra-bunch motion, and Schottky monitors for bunch-by-bunch tunes and chromaticity, among others.
They will help greatly at restart after LS1.

\section{High head-on tune shift and high pile-up}

Single bunches characterized by very high brightness were collided during dedicated MD sessions in 2011 and 2012~\cite{bib:Trad}.
First, in 2011, a possible head-on beam--beam limit was probed, with bunches characterized by $\epsilon\approx1.3\,\mu$m and $N\approx1.9\times10^{11}$\,p/bunch~\cite{bib:N29}. 
No significant losses or emittance effects were observed after having performed a tune adjustment to avoid emittance blow-up ($Q\mathrm{_H}=Q\mathrm{_V}=0.31$).
The linear head-on beam--beam parameter $\xi$ is defined as
\begin{equation}
	\xi=\frac{N r_0}{4 \pi \epsilon_n},
	\label{eqn:linpar}
\end{equation}
where $N$ is the number of protons in the bunch, $\epsilon_n$ is the normalized emittance, and $r_0=1.54\times10^{-18}$\,m is the classical proton radius.
During the 2011 experiments at injection energy, at most $\xi=0.02$/IP and $\xi=0.034$ total (for two IPs) were achieved, to be compared with the Design Report value of $\xi=0.0033$/IP for (for three head-on IPs~\cite{bib:LHC}).

Given the success of the studies at injection, bunches with similar parameters were put into collisions according to the operational cycle, so that the experiments could use such beams to study their own pile-up limitations~\cite{bib:Jaco,bib:Trad}. 
The pile-up is $\mu\approx19$ in the Design Report~\cite{bib:LHC}, but a pile-up of $\mu_{max}\approx31$ was achieved in 2011~\cite{bib:N105} and $\mu_{max}\approx70$ in 2012~\cite{bib:N10}.
The very high value achieved in 2012 was reached due to the very bright single bunches that could be produced as a result of the use of the Q20 optics in the SPS ($N=3\times10^{11}$\,p/bunch and $\epsilon=2.2\,\mu$m~\cite{bib:Papa}), and is well beyond what the experiments can handle for efficient data taking.
Even higher values would have been achieved had the beams not suffered from instabilities during the acceleration ramp and the betatron squeeze (despite the increase in chromaticity and longitudinal size).
Only one beam could be brought cleanly into collisions in the time scheduled for the study.

\subsection{Coherent Modes}

Coherent beam--beam modes, $\sigma$ and $\pi$, could be measured during the 2011 experiments with single bunches~\cite{bib:Buf3}.

It is also worth recalling that in 2010 a tune split had been used to cure instabilities, possibly from coherent modes, with single-bunch intensities of $\approx0.9\times10^{11}$\,p/bunch ($\Delta{Q\mathrm{_1}}=-0.0025$; $\Delta{Q\mathrm{_2}}=+0.0025$).
The tune split was later removed~\cite{bib:PH} when more bunches were colliding and after observing that the lifetime of one beam was significantly worse than that of the other beam (the worse lifetime being for the beam with reduced tune).

\section{Scans of the crossing angle}

In successive MD sessions~\cite{bib:Herr}, the machine settings were changed starting from the nominal configuration by reducing the crossing angle in steps until losses or lifetime reduction were observed.
This allowed the separation that corresponded to the onset of beam losses to be recorded.
Bunch-by-bunch differences depending on the number of LR interactions were highlighted (PACMAN effects), with a higher number of LR interactions leading to higher integrated losses, starting at a larger separation.
These experiments were repeated for different $\beta^{*}$ and bunch intensities; the different machine settings and beam parameters in each experiment are shown in Table~\ref{tab:scans}.
The results were used to confirm simulations~\cite{bib:Kalt} and to predict the required separation for different scenarios that might possibly be used in future operation.
It has been proven that this is a dynamic aperture effect, as no effects of the scans were observed on the emittance evolution, and as the losses recovered if a sufficient crossing angle was restored.

\begin{table}[t]
	\centering
	\caption{Machine Settings and Beam Parameters for Crossing Angle Scans ($\alpha$ is the Half Crossing Angle; $\epsilon$ is the Transverse Emittance; $\Delta$t is the Bunch Spacing; E is the Beam Energy)}
	\begin{tabular}{ccccccc}
		\toprule
		\ {\textbf{$\beta^{*}$}}	& {\textbf{$\alpha$}}	& {\textbf{Intensity}} 		& {\textbf{$\epsilon$}} & {\textbf{$\Delta$t}}& {\textbf{E}} \\
		\ \textbf{[m]} 		& \textbf{[$\mu$rad]} 		& \textbf{[$10^{11}$\,p/bu.]} 	& \textbf{[$\mu$m]} 		& \textbf{[ns]} 		& \textbf{[TeV]} \\
		\toprule
				 1.5		& 120	& 1.2		& 2--2.5	& 50		& 3.5		\\  \midrule
				 0.6		& 145	& 1.6		& 2--2.5	& 50		& 4		\\  \midrule
				 0.6		& 145	& 1.2		& 2--2.5	& 50		& 4		\\  \midrule
				 1		& 145	& 1.0		& 3.1		& 25		& 4	 	\\
		\bottomrule
	\end{tabular}
	\label{tab:scans}
\end{table}

\begin{figure}[t]
\centering
\includegraphics[angle=0, width=75mm, height=44mm]{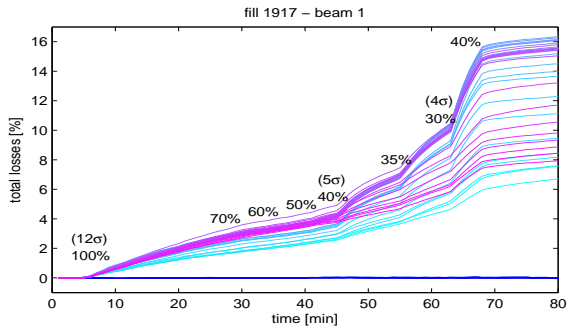}
\caption{Bunch losses versus time for Beam 1; blue curves for non-colliding bunches, and cyan to magenta for the 36 bunches in the 50\,ns spaced bunch train. The separation is indicated in the plot as a percentage of the initial crossing angle, or in the number of $\sigma$.}
\label{fig:LRlosses}
\end{figure}

\begin{figure}[b]
\centering
\includegraphics[angle=0, width=75mm, height=35mm]{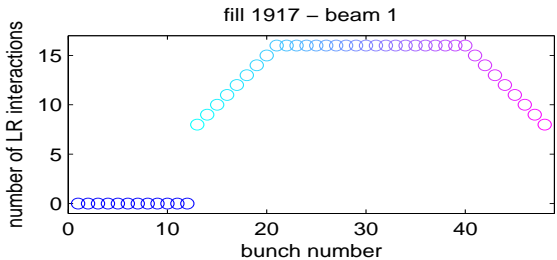}
\caption{The number of LR encounters per bunch: in blue, the 12 non-colliding bunches; in cyan fading to magenta, the 36 50\,ns spaced bunches.}
\label{fig:nLR}
\end{figure}

As an example, Fig.~\ref{fig:LRlosses} shows the losses in the case of Beam 1 for the first scan in Table~\ref{tab:scans}, when the crossing angle in IP1 was reduced from 120\,$\mu$rad, or 100\%, to a minimum of 40\% (corresponding to 4\,$\sigma$ beam separation~\cite{bib:N58}).
It can be seen that the onset of strong losses is between 4 and 5~$\sigma$ separation, depending on the number of LR interactions experienced by the bunch (shown in Fig.~\ref{fig:nLR}).

The scans served as evidence for the effectiveness of the alternate crossing scheme, since when scanning IP5 after IP1, the lifetime seemed best when the separation and the crossing angles were equal for the two IPs (we recall that the crossing plane is vertical in IP1 and horizontal in IP5, to compensate for 
first-order LR effects).
A dependence on the number of head-on collisions was also shown.

A scan was also performed for 25\,ns spaced beams -- that is, with twice the number of LR 
interactions -- as it was expected that a bigger separation will be needed, and the information will be useful in deciding the settings for future operation.
An asymmetry between Beams 1 and 2 was observed but is not yet fully understood (it is possibly related to electron cloud effects).

\section{Orbit effects}

It has been predicted that PACMAN bunches will have different orbits due to LR collisions, and a fully self-consistent treatment was developed to compute those different orbits~\cite{bib:WH}.
The LHC orbit measurement cannot resolve these effects, but the ATLAS vertex centroid 
measurement~\cite{bib:Koza,bib:Bart} was used to qualitatively verify the 
agreement~\cite{bib:N58,bib:N21,bib:Scha}.

\subsection{Missing LR Deflection}

The beam dump of a single beam in collisions leads to a transient effect due to missing LR deflections, resulting in a single-turn trajectory perturbation of the other beam.
An end-of-fill test was performed with 72 25\,ns spaced bunches ($\approx1.1\times10^{11}$\,p/bunch, $\approx65\,\mu\mathrm{rad}$ half crossing angle~\cite{bib:Baer}).
The horizontal perturbation of the Beam 1 orbit in the arc is $\approx230$\,mm = $0.6\,\sigma_\mathrm{nom}$ (with $\sigma_\mathrm{nom}=\,3.5\,\mu\mathrm{m}$). 
This leads to beam losses above dump thresholds with the physics beam. 
The effect was observed on beam losses throughout 2012.

\section{Conclusions}

The operation and performance of the LHC are strongly influenced by beam--beam effects, which, already in these first years of physics production, have driven the choice of beam parameters, machine settings, and filling schemes, so to improve performance and mitigate instabilities.
A list of observations from routine operation and dedicated studies has been presented in this paper to give an overview of the extent to which beam--beam related effects have shaped LHC operation for 
high-intensity proton physics.

\section{Acknowledgements}

The authors would like to acknowledge the operation teams of the LHC and its injectors for their observations during beam operation and their help and support during dedicated studies.
The LHC experiments are also acknowledged for providing luminosity and other beam-related data in a productive and collaborative way.

\end{document}